# Gyrokinetics investigations of an I-mode pedestal on Alcator C-Mod

Xing Liu[1], M. Kotschenreuther[1], D.R. Hatch[1], S.M. Mahajan[1], J.W. Hughes[2], A.E. Hubbard[2]

[1]University of Texas at Austin, Austin, TX 78757 USA

[2]Plasma Science and Fusion Center, Massachusetts Institute of Technology, Cambridge, MA 02139, USA

Email: mtk@austin.utexas.edu

## I-mode introduction

I-mode is a promising regime for burning plasmas. It has an edge transport barrier that gives it a high energy confinement time, as in conventional H-modes. But it has several characteristics that are superior to a conventional H-mode: 1) It avoids damaging Edge Localized Modes (ELMs) that are very problematic for ITER and proposed fusion reactors 2) It avoids the build-up of impurities in the plasma that would eventually become unacceptable. It is important to understand the physical mechanisms that operate in I-modes that give it these advantages over H-modes. The I-mode has been observed on several tokamaks, including Alcator C-Mod[1], and ASDEX-Upgrade[2] and DIII-D[3],. The C-Mod tokamak has characterized this mode especially well.

The different character of the I-mode apparently stems from the different transport processes that operate in the pedestal. The energy and particle transport seems to prevent pedestal pressure from reaching an ideal-MHD stability boundary, so that ELMs are avoided. The particle transport seems to prevent impurity and main-species density build-up. To understand the I-mode better, we have used the gyrokinetic code GENE[4,5,] to examine the instabilities and transport in the pedestal of a particular high performance I-mode discharge on C-Mod: shot number 1120907032 at 1.0s.

The C-Mod I-mode pedestals are observed to have a unique fluctuation called a Weakly Coherent Mode (WCM). This is suspected of being at least partially responsible for the unique characteristics of the I-modes. Hence, one of the important goals of our investigation is to find instabilities in the gyrokinetic simulations that correspond to the observed WCM. Another goal is to clarify the physical mechanisms of the fluctuation.

# Global linear simulation results

We used GENE to perform "global" simulations of the pedestal, which means that the full radial profile variations in the pedestal region were included. The full poloidal extent is also simulated assuming equilibrium temperatures and densities are constant along a flux surface with a full variation of magnetic field. In a global linear run, a single toroidal mode number is simulated, while nonlinear runs include multiple toroidal numbers (details given below). The simulation box was large enough to include the entire pedestal, as indicated in Fig 1 and Fig 2. Here the fitted pressure profile, and $T_e$, $T_i$, $n_e$ profiles from experiment are shown, and the vertical lines indicate the boundaries of the simulation box. We also use the measured radial electric field (Fig 3). For all GENE global runs, a "buffer" zone is included near the boundaries, where extra damping is added to ensure good numerical behavior. The radial coordinate we use in the simulations and in this paper is the normalized square root of toroidal flux ($\rho_t = \sqrt{\frac{\Psi_t}{\pi B_0}}$).

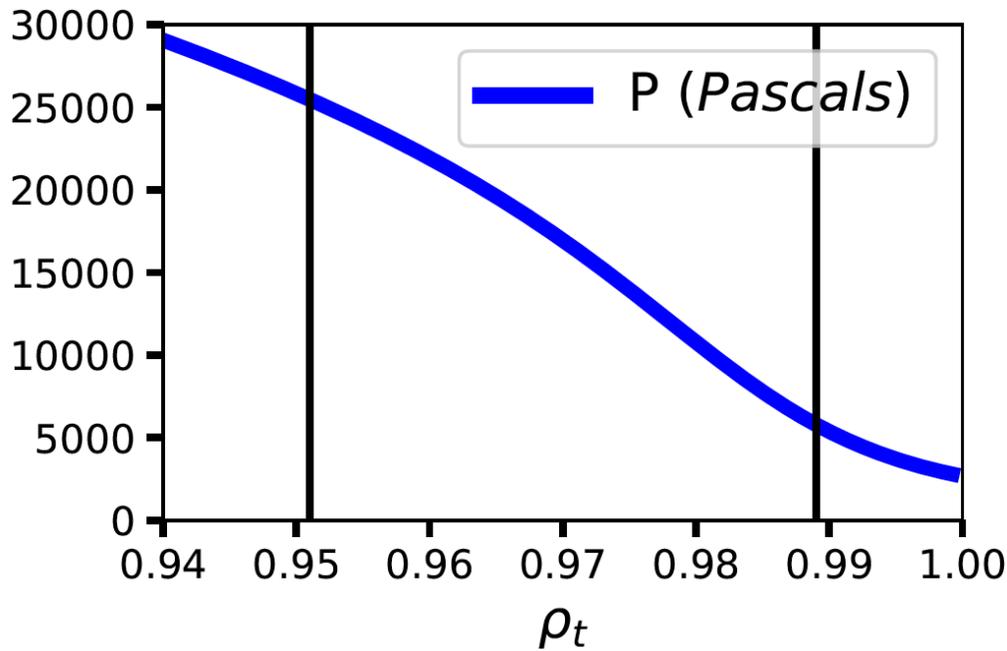

Figure 1: Total pressure profile (P) and simulation box (two vertical lines are boundaries). Simulation box chosen to include the pedestal.

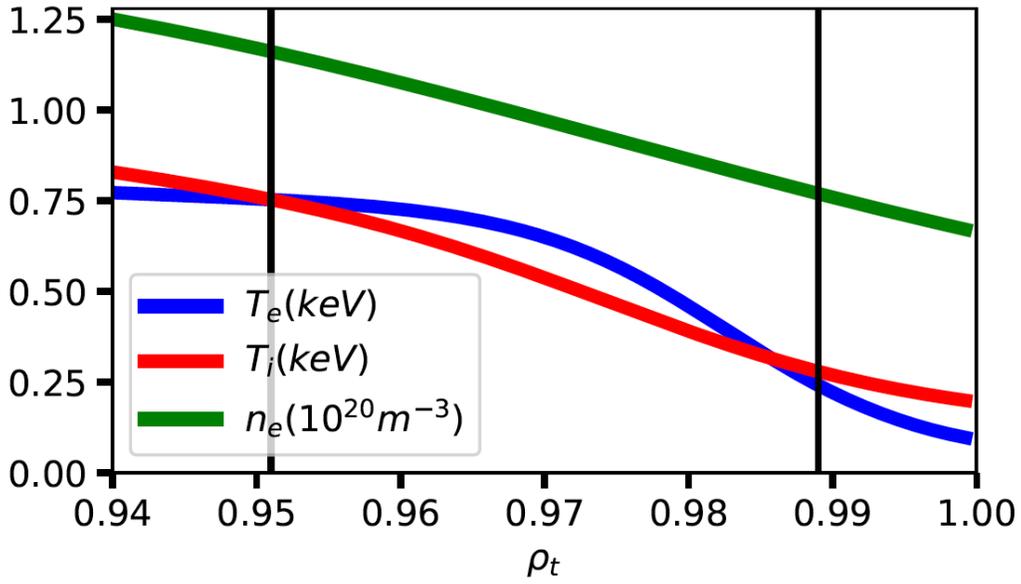

Figure 2: $T_e$, $T_i$, $n_e$ profiles and simulation box. Black vertical lines indicate the boundaries of simulation box.

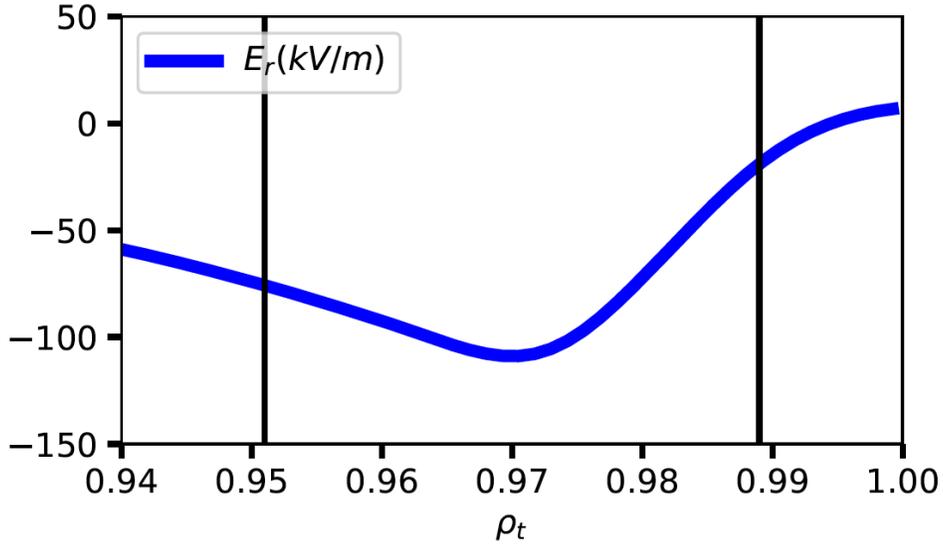

Figure 3: $E_r$ profile and simulation box. Black vertical lines indicate the boundaries of simulation box.

We conduct simulations based on the experimental profiles of $T_e$, $n_e$ from Thomson scattering and $T_i$, $E_r$ from gas puff based charge exchange recombination spectroscopy. We use an MHD equilibrium reconstructed from EFIT using the measured pressure profiles. The impurity $Z_{eff}$ is estimated to be 2.8, and the averaged impurity charge $Z_{avg}$ is 10 (a combination of Boron and metal), and these are assumed constant across the pedestal in order to infer $n_i$ and $n_z$ profiles. The most unstable mode found in GENE global simulations for the toroidal wavenumbers where peak WCMs are observed, are shown in Table 1.

| Toroidal mode number (n) | Growth rate $\gamma(c_s/a)$ | Frequency in lab frame (kHz) | Averaged ExB Doppler shift (kHz) | $\chi_i/\chi_e$ | $D_e/\chi_i$ | $D_z/\chi_i$ | $Q_{es}/Q_{em}$ | $<E_\parallel>$ | $E_{\parallel,em}/E_{\parallel,es}$ |
|---|---|---|---|---|---|---|---|---|---|
| 15 | 0.079 | -271.67 | -255.74 | 2.65 | 0.034 | 1.4 | -25.79 | 0.96 | 0.029 |
| 16 | 0.081 | -291.06 | -273.03 | 2.58 | 0.046 | 1.41 | -26.33 | 0.97 | 0.025 |
| 17 | 0.082 | -310.13 | -290.32 | 2.5 | 0.055 | 1.42 | -24.3 | 0.97 | 0.024 |
| 18 | 0.085 | -329.21 | -308.14 | 2.41 | 0.064 | 1.42 | -22.48 | 0.98 | 0.018 |
| 19 | 0.087 | -348.49 | -325.96 | 2.34 | 0.077 | 1.43 | -21.43 | 0.97 | 0.023 |

Table 1: Mode frequency in lab frame of experimental toroidal mode numbers (15—19) are in the peak WCM frequency range (-300– -400 kHz). Negative frequency values indicate electron diamagnetic direction. Most of the mode frequency is from the ExB Doppler shift frequency. The mode is essentially electrostatic. $\chi_i/\chi_e$ (>1) is consistent with ITG transport fingerprints.

The mode frequencies in the lab frame, where WCM density and magnetic fluctuations are measured to peak, are found to be in the -300 — -400 kHz range. (Negative frequency values indicate electron diamagnetic direction.) The frequencies are fairly close to the ExB Doppler shift, arising from the radial electric field $E_r$-well in the pedestal, at the radial location where the mode peaks. This is consistent with ITG, since in the plasma frame, the frequency of ITG is, typically, a small fraction of ion diamagnetic frequency ($\omega_i*$). It could also be possibly consistent with resistive ballooning mode or TEM.

The ion thermal transport is the dominant transport channel affected by this mode. The ratio of the induced ion to electron thermal diffusivity ($\chi_i/\chi_e$) is around 2.5. The mode is essentially electrostatic ($<E_\parallel>\sim 1$ and $E_{\parallel,em}/E_{\parallel,es}<<1$). The ratio of the diffusivity in different channels, which we call the transport "fingerprints", are typical ITG mode fingerprints. The physical reason for why these modes have low $D_e/\chi_i$ in this regime will be discussed later. In contrast, MHD-like modes (which are electromagnetic, and have an inductive $A_\parallel$ field to cancel $\nabla_\parallel \varphi$), generally have similar diffusivities of all quantities (i.e., $\chi_e/\chi_i \sim 1$, $D_e/\chi_i \sim 2/3$, and $D_Z/\chi_i \sim 2/3$)[6]. Note that we compute the "effective" particle diffusivity by dividing the quasilinear particle flux by the density gradient. We don't split the diffusive term (D dn/dx) and the inward pinch term (Vn) in the particle fluxes because there is no clear procedure to do this for instability caused transport. Therefore, if inward pinch dominates, we would have a negative value for the effective particle diffusivity.

### ION TEMPERATURE GRADIENT SCALE LENGTH (A/$L_{T_I}$) SCAN

In order to probe further into the nature of the mode, we vary the $T_i$ profile while keeping the other profiles the same. New $T_i$ profiles are made according to the formula,

$$T_i(\rho_t) = T_i(\rho_t = 0.97) \cdot \left[ \frac{T_i(\rho_t)}{T_i(\rho_t = 0.97)} \right]^\alpha$$

where $\alpha$ = 0.8, 0.9, 1.1 or 1.2. The ion temperature is unchanged at the middle of the pedestal ($\rho_t = 0.97$). In this way, the normalized ion T gradient scale length, a/$L_{T_i}$, is varied by a factor of $\alpha$. Global linear simulations (Table 2) find the growth rate of this mode to increase as $T_i$ gets steeper and to become stable when it becomes less steep (i.e, when a/$L_{T_i}$ is 0.8 times the observed value, the growth rate drops and the dominant instability goes into an ETG mode).

| a / $L_{T_i}$ factor | Growth rate $\gamma (c_S/a)$ | Frequency in lab frame (kHz) | Averaged ExB Doppler shift (kHz) | $\chi_i /\chi_e$ | $D_e/\chi_i$ | $D_Z /\chi_i$ | $Q_{es}/ Q_{em}$ | $<E_\parallel>$ | $E_{\parallel,em}/E_{\parallel,es}$ |
|---|---|---|---|---|---|---|---|---|---|
| 1.2 | 0.165 | -320.93 | -305.00 | 2.9 | 0.014 | 1.16 | -17.27 | 0.98 | 0.02 |

| 1.1 | 0.127 | -325.02 | -307.09 | 2.67 | 0.039 | 1.28 | -18.52 | 0.98 | 0.019 |
|-----|-------|---------|---------|------|-------|------|--------|------|-------|
| 1 | 0.085 | -329.21 | -308.14 | 2.41 | 0.064 | 1.42 | -22.48 | 0.98 | 0.018 |
| 0.9 | 0.033 | -333.61 | -309.19 | 2.16 | 0.1 | 1.57 | -42.45 | 0.98 | 0.016 |
| 0.8 | 0.032 | -233.83 | -246.30 | 0.44 | -0.1 | 1.86 | -19.47 | 0.94 | 0.055 |

Table 2: Ion temperature gradient scale length variation result.

Thus, the ion temperature gradient is a significant drive for this mode.

The simulations find that the poloidal structure of this mode (Fig 4) is unlike most modes found in the core; it doesn't peak at the outboard midplane (z = 0 is outboard midplane). Radially, the mode peaks at the outer region of the $E_r$-well (bottom of the $E_r$-well is at $\rho_t = 0.97$) where $\eta_i$ drive is large enough to overcome shear suppression ($\gamma_{ExB}$) (see Fig 5 for $\eta_i$ profile and Fig 6 for $\gamma_{ExB}$ profile). Closer to the separatrix, shear suppression becomes too large for the mode to grow, while near the core, $\eta_i$ is not strong enough to destabilize the modes. A detailed study of the scaling of such turbulence with ExB shear is provided in Ref. 7. Eigenfunctions from global linear simulation (a typical eigenfunction of $\varphi$ shown in Fig 4) are consistent with experimental observation of the mode's location.[8]

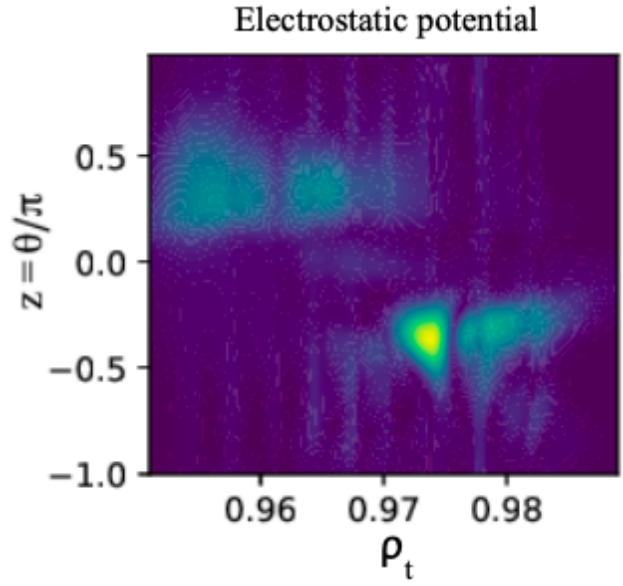

Figure 4: Mode structure of the electrostatic potential ($\varphi$) in a global simulation for the base case (which is typical). Here z is poloidal angle $\theta$ divided by $\pi$, and $\rho_t$ is the radial coordinate. As is seen from the graph, the mode peaks considerably away from the outboard midplane (z = 0).

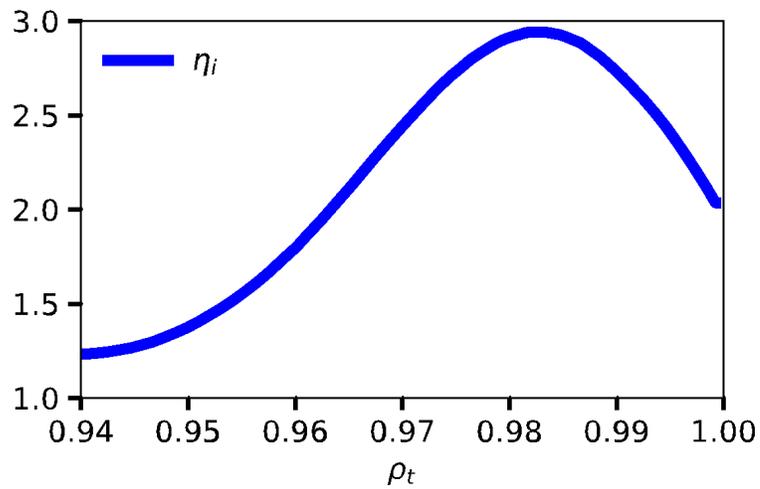

Figure 5: $\eta_i$ profile in the pedestal region.

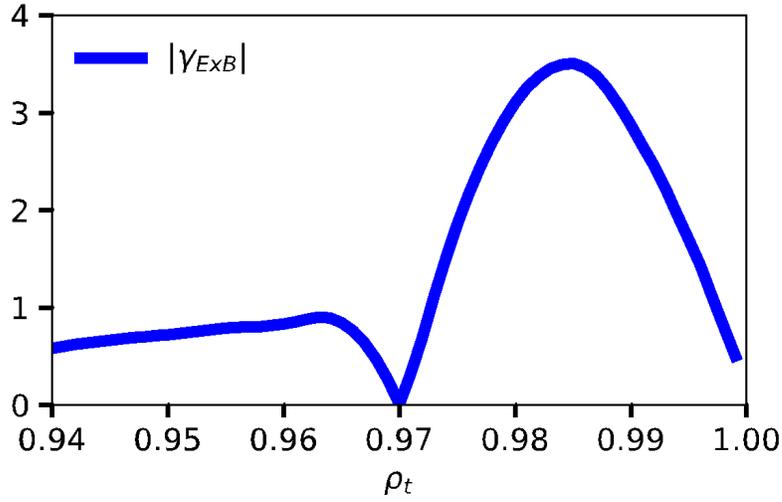

Figure 6: Absolute value of $\gamma_{E \times B}$ from experimental $E_r$ profile in the pedestal region.

## DENSITY GRADIENT SCALE LENGTH (A/L$_{NE}$) SCAN

The next parameter scanned is the density gradient. Density profiles of all three simulated species are varied together to satisfy quasi-neutrality, and constant $Z_{eff}$ (= 2.8). We also keep total pressure unchanged from the experimental profiles by modifying the temperature profiles accordingly. With total pressure the same, the pressure gradient drive of MHD-like modes is kept constant. In Table 3, $f_{pl}$ is the frequency of the mode in the plasma frame in normalized units.

| a/L$_{ne}$ factor | Growth rate $\gamma(c_S/a)$ | Frequency in plasma frame $f_{pl}(c_S/a)$ | $\chi_i/\chi_e$ | $D_e/\chi_i$ | $D_z/\chi_i$ | $Q_{es}/Q_{em}$ | $<k_z>$ ($|\varphi|^2$ averaged) |
|---|---|---|---|---|---|---|---|
| 0.8 | 0.181 | -0.05 | 2.73 | -0.07 | 0.99 | -16.85 | 0.24 |
| 0.9 | 0.136 | -0.13 | 2.62 | 0.005 | 1.19 | -18.08 | 0.25 |
| 1 | 0.085 | -0.20 | 2.41 | 0.064 | 1.42 | -22.48 | 0.25 |

| 1.1 | 0.017 | -0.26 | 1.97 | 0.14 | 1.59 | -62.81 | 0.25 |
|-----|-------|-------|------|------|------|--------|------|
| 1.2 | -0.004 | -0.35 | 2.55 | 0.1 | 2.21 | 128.2 | 0.31 |

Table 3: Density gradient scale length variation indicates the mode is destabilized by lower density gradient when total pressure gradient is kept constant. Frequency in plasma frame is the frequency of the mode in the lab frame subtracted by the eigenfunction averaged ExB Doppler shift frequency.

As expected, the growth rate decreases when the density gradient is increased (Table 3); the mode becomes stable when $a/L_{ne}$ is 20% above the experimental value. This provides additional evidence that we are dealing with an ITG mode driven by $\eta_i$; a pressure gradient driven MHD-like mode would not be stabilized. After computing the mode wavenumber $k_z$ (along the magnetic field line) from the eigenfunction $\varphi$, we find that the mode frequency in the plasma frame, is close to $k_z v_{th,i}$ ($k_z v_{th,i} / f_{pl} <\sim 1$), *indicating that ion thermal resonance is important for this instability*. A kinetic treatment is therefore necessary in identifying and explaining this mode. This could be the reason why the fluid treatment of Liu et. al. did not find this mode[9]. Passing electrons, however, could be considered adiabatic since $k_z v_{th,e}$ is much larger than the mode frequency ($k_z v_{th,e} / f_{pl} > 1$). Adiabatic electron response would explain why $D_e /\chi_i$ is small for this mode.

**COLLISIONALITY SCAN**

Experiments found that I-mode discharges have low collisionality pedestals ($\nu^*_{95} \sim 0.1$) compared to H-mode with the same pedestal temperature[8] and WCM signal is clearest on diagnostics in low $\nu^*$ pedestals[9]. To study the effects of pedestal collisionality on the mode's growth rate, new profiles are created by multiplying the temperature profile by a given factor and dividing the density profile by the same factor. In this way, collisionality is modified while total pressure is kept constant and consistent with the experimental pressure.

| Collisionality factor | Growth rate $\gamma(c_s/a)$ | Frequency $\omega(c_s/a)$ | Frequency in plasma frame $f_{pl}(c_s/a)$ | $<\nu_{ei}>$ | $\nu_{eff}/|\omega_{pl}|$ | $<\omega_{bounce,e}>$ | $<\nu^*>$ |
|---|---|---|---|---|---|---|---|
| 0.1 | 0.109 | -1.216 | 0.396 | 0.272 | 1.69 | 6.434 | 0.108 |
| 0.8 | 0.09 | -2.977 | -0.177 | 1.297 | 16.63 | 4.441 | 0.744 |
| 1 | 0.085 | -3.141 | -0.202 | 1.630 | 18.94 | 4.329 | 0.959 |
| 1.2 | 0.081 | -3.278 | -0.223 | 1.974 | 21.19 | 4.256 | 1.181 |
| 2 | 0.077 | -3.593 | -0.281 | 3.03 | 26.44 | 4.148 | 1.859 |
| 4 | 0.074 | -4.077 | -0.34 | 5.83 | 42.82 | 4.099 | 3.621 |

Table 4: Growth rate is not very sensitive to the collisionality. Collisional detrapping of the electrons is important ($\nu_{eff}/|\omega_{pl}| >1$), making trapped electron effect weak.

In Table 4, the electron bounce frequency is $\omega_{bounce,e} = \sqrt{\epsilon}\frac{v_{th,e}}{qR}$ and $<...>$ denotes weighted average by eigenfunction $\varphi$. $|\omega_{pl}|$ is the absolute value of the complex frequency of the mode in the plasma frame. Note that since the mode peaks toward the separatrix, the mode averaged $\nu^*$ is significantly higher than $\nu^*_{95}$.

By changing collisionality from the experimentally observed value, we found that the modes tend to become slightly more stable at higher collisionality. This is in contrast with the resistive ballooning mode that has a higher growth rate at higher collisionality. To quantify the importance of collisions for the trapped electrons, we normalize $\nu_{eff}$ ($\nu_{eff} = \nu_{ei}/\epsilon$, is the collisional electron de-trapping rate, as in neoclassical theory) by the absolute value of the complex frequency of the mode in the plasma frame, $\nu_{eff}/|\omega_{pl}|$. Since this value for the experimental profile is much larger than 1 (~ 20), it means that collisions detrap electrons much more frequently than mode frequency of this mode. Hence the trapped electron effect is weak. Note that the mode frequency in the plasma frame is much smaller than the electron bounce frequency. Even though $\nu^*$ is on the order of 1, the relevant definition of collisionality for this mode, $\nu_{eff}/|\omega_{pl}|$, shows that this regime is too collisional for the trapped electrons to be a dominant effect on this instability.

This also indicates that the non-adiabatic trapped electron effects are small. Since both passing and trapped electron non-adiabaticity is weak, it follows that the electron particle transport is weak for this mode, i.e., $D_e/\chi_i$ is small.

Hence this is a slab-like ITG where curvature drive and trapped electron drive are not of primary importance.

**IMPURITY DENSITY GRADIENT ($A/L_{NZ}$) SCAN**

Since impurity density profile (difficult to measure in the experiment) is known to affect the instability of ITG, we run simulations to probe the sensitivity of this variable and look for the most probable impurity density $n_z$ profile in the steady state. (An impurity mode has been proposed as an explanation for the WCM.[12]) In our scan of the gradient of the impurity density profile, we go from one extreme to the other – from a flat profile to one slightly steeper than the electron density profile. The electron density profile is fixed at the measured value, and main ion density profile is adjusted accordingly to satisfy quasi-neutrality. The results in Table 5 show that low impurity density gradient destabilizes the mode. (Note that, since $n_e$ decreases as the LCFS is approached, but $n_z$ is nearly constant, $Z_{eff}$ is rising toward the LCFS). This destabilizing trend is consistent with the effect of impurities on ITG[13].

| $a/L_{nz}$ factor | Toroidal mode number (n) | Growth rate $\gamma(c_s/a)$ | $\chi_i/\chi_e$ | $D_e/\chi_i$ | $D_z/\chi_i$ | $Q_{es}/Q_{em}$ | $<E_\parallel>$ | $E_{\parallel,em}/E_{\parallel,es}$ |
|---|---|---|---|---|---|---|---|---|
| 0 | 18 | 0.168 | 2.56 | 0.01 | $-\infty$ | -18.67 | 0.98 | 0.019 |
| 0.2 | 18 | 0.154 | 2.55 | 0.016 | -0.9 | -19.1 | 0.98 | 0.018 |
| 0.3 | 18 | 0.147 | 2.54 | 0.021 | 0.066 | -19.4 | 0.98 | 0.018 |
| 0.4 | 18 | 0.139 | 2.54 | 0.027 | 0.55 | -19.74 | 0.98 | 0.018 |
| 0.8 | 18 | 0.103 | 2.47 | 0.052 | 1.28 | -21.46 | 0.98 | 0.018 |
| 1 | 18 | 0.085 | 2.41 | 0.064 | 1.42 | -22.48 | 0.98 | 0.018 |
| 1.2 | 18 | 0.065 | 2.35 | 0.076 | 1.52 | -23.5 | 0.98 | 0.018 |

Table 5: Impurity density gradient scale length variation shows that low impurity density gradient destabilizes the mode.

The impurity particle diffusivity is found to be very sensitive to the impurity density gradient. When $a/L_{nz}$ is equal to or slightly larger than $a/L_{ne}$, the ratio of impurity particle diffusivity to ion thermal diffusivity ($D_Z/\chi_i$) could be as high as $\sim 1.5$. This implies that this mode is an effective channel in expelling impurities and flattening impurity density gradient if impurities accumulate inside the pedestal. The value $a/L_{nz}$ is reduced to be a small fraction of $a/L_{ne}$ to reveal the pinch term, since the effective impurity particle diffusivity becomes negative, indicating a weak inward impurity pinch caused by this mode. In a steady-state discharge such as I-mode, since there's no significant impurity particle source inside the plasma, the flux of impurity particles must be close to zero. Therefore, based on the quasilinear particle flux from these linear simulations, the most probable impurity density profile, based on the $D_Z/\chi_i$ produced by the mode, is the one with $a/L_{nz} \sim 0.3 * a/L_{ne}$, where $a/L_{ne}$ is the observed electron density profile scale length. (Recall as mentioned above, we model impurity as a single composite species which reflects average of all impurities.)

SUMMARY OF CONCLUSIONS FOR THE DENSITY TRANSPORT

*The results above are fully consistent with the observed electron density profile.* The density gradient at which $D_e/\chi_i$ goes to zero is with the density profile modification that makes $a/L_{ne} \sim 0.9$ times the observed value, i.e., very close to the experimental value in steady-state. The results above can be interpreted as showing that the low particle loss for this mode is sustained by a small outward diffusion and small inward pinch. This is similar to conclusions from gyrokinetic simulations ITG transport in the JET pedestal[14] and consistent with the small particle source inferred for I-mode as we will discuss below.

We found that, as the impurity density gradient ($a/L_{nz}$) and ion temperature gradient ($a/L_{Ti}$) are varied, this mode has a low ratio of electron particle diffusivity to ion thermal diffusivity ($D_e/\chi_i$). This is to be expected for this mode since particle transport only appears when electrons are non-adiabatic. As is described above, the passing electrons are adiabatic because $k_z\, v_{th,e} \gg f_{pl}$. Trapped electron effects are weak because they are detrapped by collisions ($v_{eff}/|\omega_{pl}| > 1$). In addition to that, $\varphi$ doesn't peak at outboard midplane where electrons were trapped, which further reduces the coupling of this mode to trapped electrons. *Low particle transport is therefore unavoidable consequence of the basic physics of this mode.*

We now examine the consistency of this observation, which is at first sight counterintuitive for the I-mode regime with mainly a thermal transport barrier, with experimental inferences.

**DIFFUSIVITY INFERENCES BASED ON EXPERIMENTAL PROFILES AND SOURCES**

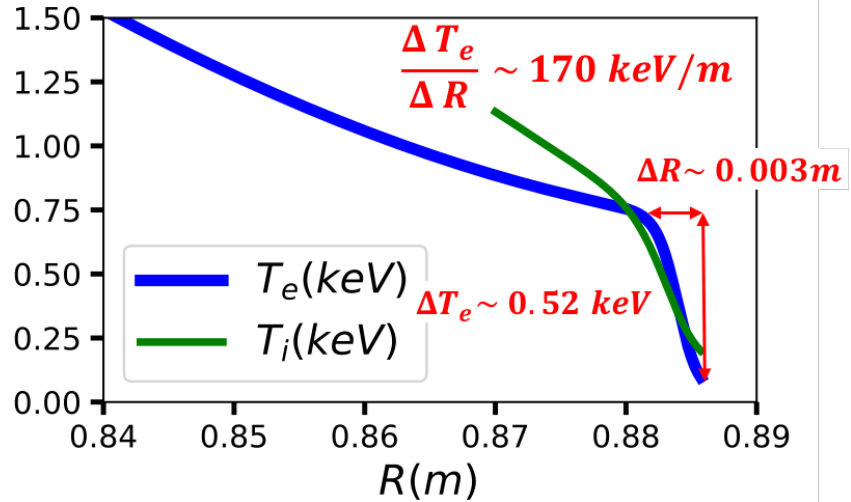

Figure 7: Electron and ion temperature profiles in the pedestal region.

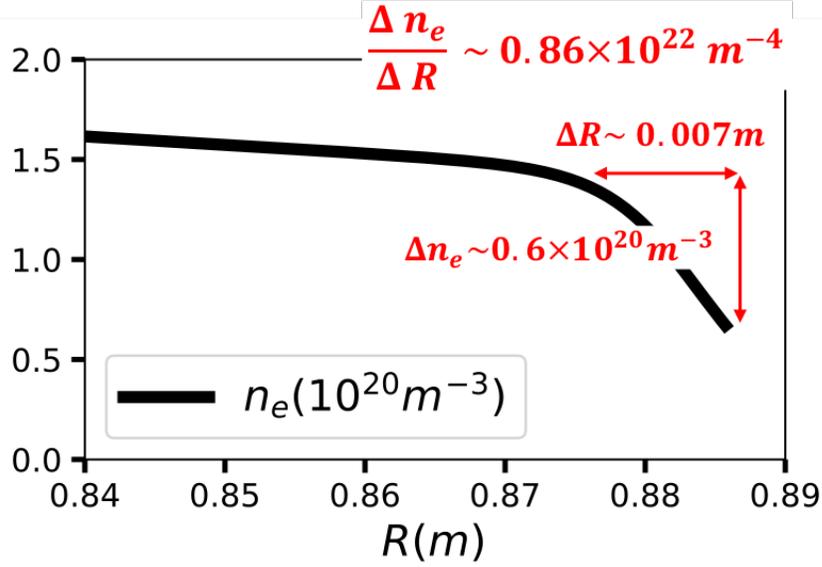

Figure 8: Electron density profile in the pedestal region.

We estimate the pedestal thermal diffusivity using the energy flux through the last closed flux surface (LCFS), and the electron temperature gradient in the pedestal. The energy flux through the pedestal is estimated using $P_{Net} = P_{ICRF} + P_{Ohmic} - P_{Radiation} \sim$ 4 MW and surface area ($A \sim 7\ m^2$):

$$Q \sim \frac{P_{heating}}{A_{LCFS}} = \frac{4\ \times\ 10^6\ W}{7\ m^2} \sim 0.57\ \times\ 10^6\ Wm^{-2}$$

The electron density in the middle of the pedestal and the average temperature gradient are, respectively, estimated as $n_e \sim 1.0 \times 10^{20}\ m^{-3}$, and $\frac{\Delta T_e}{\Delta R} \sim 170\ keVm^{-1}$, $\frac{\Delta T_i}{\Delta R} \sim \frac{1}{2}\ \frac{\Delta T_e}{\Delta R}$ (See the Fig 7 and Fig 8). If the electron thermal diffusivity dominates, the thermal diffusivity is found to be $\chi^{eff} = 0.2\ m^2 s^{-1}$; if the electron and ion thermal diffusivities are the same, the thermal diffusivity is found to be $\chi^{eff} = 0.13\ m^2 s^{-1}$.

If $\chi_e \gg \chi_i$,

$$Q = \chi^{eff}\ n\ \frac{dT_e}{dR} \sim \chi^{eff}\ n\ \frac{\Delta T_e}{\Delta R}$$

$$\chi^{eff} = \frac{Q}{n\frac{\Delta T_e}{\Delta R}} = \frac{0.57 \times 10^6\ Wm^{-2}}{1.0 \times 10^{20}\ m^{-3}\ \times 170\ keVm^{-1}} \sim 0.2\ m^2 s^{-1}$$

If $\chi_e = \chi_i$,

$$Q = \chi^{eff} \, n \left( \frac{dT_e}{dR} + \frac{dT_i}{dR} \right) \sim \chi^{eff} \, n \left( \frac{\Delta T_e}{\Delta R} + \frac{\Delta T_i}{\Delta R} \right)$$

$$\chi^{eff} = \frac{Q}{n \left( \frac{\Delta T_e}{\Delta R} + \frac{\Delta T_i}{\Delta R} \right)} = \frac{0.57 \times 10^6 \, Wm^{-2}}{1.0 \times 10^{20} \, m^{-3} \, \times 1.5 \times 170 \, keV m^{-1}} \sim 0.13 \, m^2 s^{-1}$$

We estimate the effective electron particle diffusivity ($D_e^{eff}$) using the particle flux through the LCFS, and the electron density scale length in the pedestal region. The shot we analyze here has plasma current of Ip = 1.2 MA. Then, from Table 5.1 in Dominguez Ph.D thesis[15], the electron particle flux through the LCFS, $\Gamma_{LCFS}$ is calculated, based on measurements, to be $\Gamma_{LCFS} = 1.2 \times 10^{20} \sim 1.5 \times 10^{20} \, m^{-2}s^{-1}$. The shot we analyze here has somewhat more gas puffing than in the Dominguez data set, and his experimental value is only for the midplane particle flux. Nonetheless, we use his estimate as the best that is currently available for this C-Mod I-mode since the diagnostic employed in his work is not available. We estimate the average density gradient to be $0.86 \times 10^{22} \, m^{-4}$ (See Fig 8). The effective electron particle diffusivity ($D_e^{eff}$), found to be in the range $D_e^{eff} = 0.014 \sim 0.017 \, m^2 s^{-1}$, is over an order of magnitude smaller than the effective thermal diffusivity,

$$\Gamma_{LCFS} = D_e^{eff} \frac{dn_e}{dR} \sim D_e^{eff} \frac{\Delta n_e}{\Delta R}$$

$$D_e^{eff} = \frac{\Gamma_{LCFS}}{\frac{\Delta n_e}{\Delta R}} = \frac{1.2 \times 10^{20} \sim 1.5 \times 10^{20} \, m^{-2}s^{-1}}{0.86 \times 10^{22} \, m^{-4}} = 0.014 \sim 0.017 \, m^2 s^{-1}$$

It should be noted that there is significant uncertainty as to the particle source in this shot; it is plausibly a few times larger. Nonetheless, the order of magnitude of the difference between $D_e^{eff}$ and $\chi^{eff}$ is at least qualitatively consistent with the basic physics of the mode we find.

Total impurity flux out from the LCFS ($\Gamma_Z$) is evaluated by dividing the total impurity particles inside the confined plasma by the impurity confinement time and the LCFS area. The former is obtained by multiplying the average impurity density in the core by the volume within the LCFS (V ~ 0.88 m³), while the impurity confinement time is estimated to be $\tau_Z$ ~ 30 ms.[13]

$$\Gamma_z = \frac{V \cdot n_{z,core}}{\tau_z \cdot A_{LCFS}}$$

Another way of computing the impurity flux is through using the pedestal diffusivity,

$$\Gamma_z = D_z \frac{dn_z}{dx} \sim D_z \frac{n_{z,ped}}{2\,w_{ped}}$$

Since the core impurity profile is typically flat, we assume that the impurities are mainly confined by the pedestal. Therefore, the two ways of calculating impurity flux should be equal,

$$\frac{V \cdot n_{z,core}}{\tau_z \cdot A_{LCFS}} = D_z \frac{n_{z,ped}}{2\,w_{ped}} \Rightarrow D_z \geq \frac{V \cdot 2\,w_{ped}}{\tau_z \cdot A_{LCFS}}$$

Here we assume $n_{z,core} \geq n_{z,ped}$ and the scale length of the impurity density profile to be same as the electron density scale length: $w_{ped} \sim 0.01$ m. We find the lower bound of $D_z^{eff} >= 0.08\ m^2 s^{-1}$. This means that the impurity particle diffusivity is several times higher than the electron particle diffusivity estimated above.

As we will see below, our nonlinear simulations reach this same conclusion. In particular, they are able to reproduce the relatively short impurity lifetime observed in laser blow-off experiments in I-modes on C-Mod.

GLOBAL SIMULATION GRID CHOICE AND CONVERGENCE TEST

Now we will explain some technical aspects of our simulation. As is indicated on Fig 1, Fig 2 and Fig 3, the radial simulation box, centered at $\rho_t = 0.97$ (location of the $E_r$-well bottom), extends between $\rho_t \sim (0.95, 0.99)$; the simulation box is about 16 gyroradii (16 $\rho_S$) wide. The actual simulation zone lies between $\rho_t \sim (0.96, 0.98)$ flanked by a buffer zone of width $\Delta\rho_t \sim 0.01$ on each side.

Global simulations are done on a simulation grid of (128, 72, 48, 32) in terms of (nx, nz, nv, nw). The desirable number of grid points in each dimension is determined by performing convergence tests, increasing resolutions in x, z, and velocity space (v and w) by 1.5 times each. The results are summarized in Table 6. Since the growth rates and frequencies found in higher resolution runs don't deviate from the original case by more than 10%, we decide to use the original resolutions to reduce computing time.

| nx | nz | nv | nw | Growth rate $\gamma(c_S/a)$ | Frequency $\omega(c_S/a)$ |
|---|---|---|---|---|---|
| 128 | 72 | 48 | 32 | 0.085 | -3.141 |
| 192 | 72 | 48 | 32 | 0.084 | -3.142 |
| 128 | 72 | 72 | 48 | 0.090 | -3.124 |
| 128 | 108 | 48 | 32 | 0.082 | -3.152 |

Table 6: Convergence test results.

## Nonlinear global simulations

Having identified the dominant instability, we turn to nonlinear global simulations to work out its nonlinear consequences, in particular, the transport caused by the ITG/Impurity mode. The idea is to compare it to the experimental input power of this shot: $P_{Net} = P_{ICRF} + P_{Ohmic} - P_{Radiation} \sim 4$ MW.

Note that these nonlinear simulations include the full kinetic dynamics of n=0 perturbations. In particular, they include zonal flows, Geodesic Acoustic Modes and other acoustic modes, and local profile modifications. They do not include multi-scale interactions of ETG modes and ion-scale modes. Simulations that include multi-scale effects are extremely computationally expensive and are left to future work.

### NUMERICAL DETAILS

Global ion-scale GENE simulations use 128 grid points in the radial direction (over a domain of 16 gyroradii), 72 gridpoints in the parallel direction (spanning poloidal angle from $-\pi$ to $\pi$), 48 gridpoints in parallel velocity (spanning the range -3.2 to 3.2 $c_s$, where $c_s$ is the sound speed), and 24 points in magnetic moment $\mu$ (spanning 0-10.1 $T_e/B_0$). The robust electron scale transport (described below) makes it difficult to resolve the dynamics in the $k_y$ coordinate; ion scale simulations exhibit non-negligible contributions to the heat flux at the high ky cutoff wavenumbers. In lieu of full multi-scale simulations (not accessible within the computational budget allocated to this work), we ran single-scale ETG simulations (described below) and single scale ion-scale simulations with various numerical setups: 16-64 toroidal mode numbers, $\Delta n$=3-6 and tests with and without 4[th] order hyper-diffusion in the $k_y$. The highest resolution simulations employed 64 toroidal mode numbers (ranging from n=0 to n=189 with $\Delta$n = 3, or alternatively, $k_y\rho_s$=0-1.89 with $k_{y,min}\rho_s$=0.03) and exhibit well-behave heat flux spectra as shown in Fig 9. The transport fluxes for these different simulations exhibited quantitative differences but no qualitative differences—i.e., all simulations support the main conclusions described below.

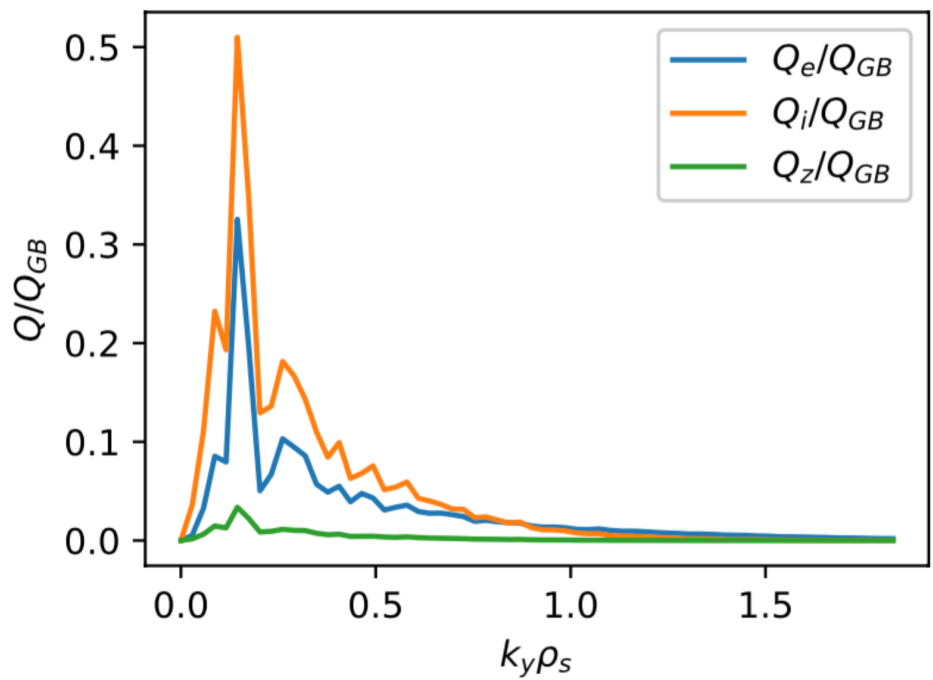

Figure 9: Spectra of electrostatic heat flux in the ion, electron, and impurity channels for a representative global ion scale nonlinear simulation.

**NONLINEAR ITG/IMPURITY THERMAL TRANSPORT**

Radial profiles of heat loss for three impurity profiles are shown in Fig 10 and Fig 11. As an upper bound, the largest amount of heat loss found in the nonlinear runs is found using a flat impurity profile (a/$L_{nz}$ = 0). It also has the highest linear growth rate in the a/$L_{nz}$ scan above. For this case, peak heat loss is ~ 0.65 MW (0.5 MW from ions, 0.15 MW from electrons). Peak heat loss for the impurity profile which is close to what we think is steady state (a/$L_{nz}$ ~ 0.4 * a/$L_{ne}$) is ~ 0.45 MW (0.35 MW from ions, 0.1 MW from electrons). For the run whose impurity profile has the same density scale length as the electron density profile, the peak heat loss is ~ 0.25 MW (0.15 MW from ions, 0.1 from electrons). Since the total net power coming into pedestal for this shot is about 4 MW, ITG/Impurity mode is an order of magnitude too low to match power balance.

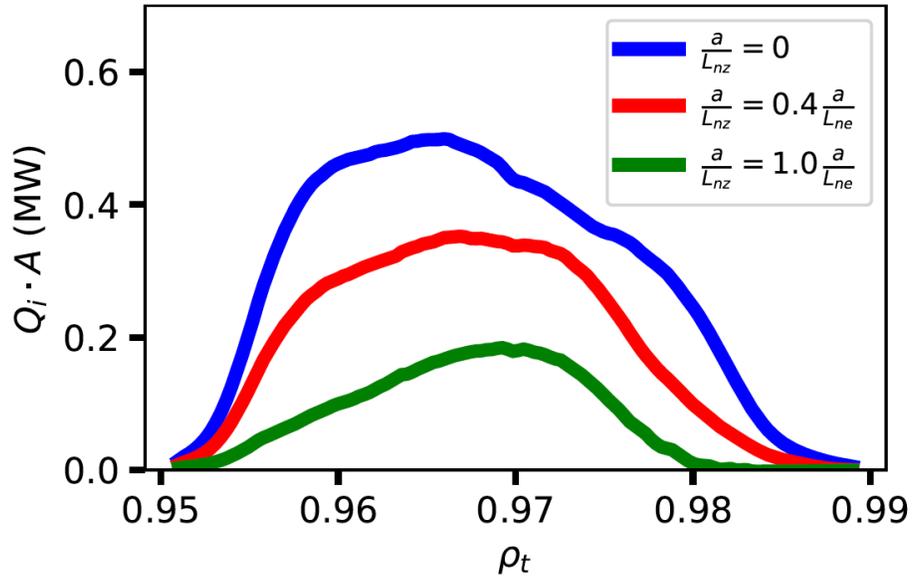

Figure 10: Radial profile of ion electrostatic heat loss for simulations using different impurity profiles.

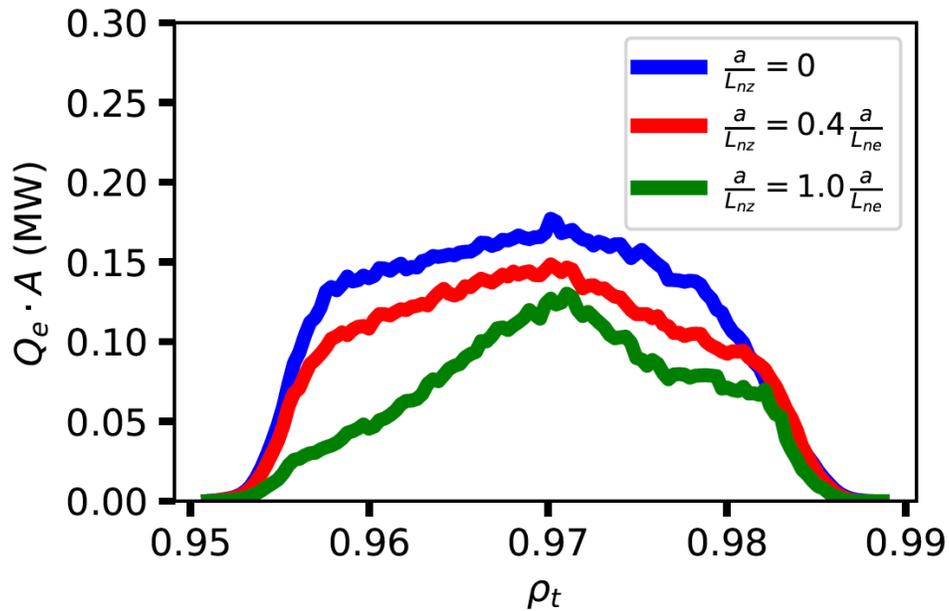

Figure 11: Radial profile of electron electrostatic heat loss for simulations using different impurity profiles.

**NONLINEAR ETG THERMAL TRANSPORT**

We run local flux tube nonlinear ETG simulations to find the heat loss from electron scale turbulence. In the nonlinear ETG simulations described here, the main ion and impurity species are assumed adiabatic and their effect enters into the simulation in the parameter $\tau$, $\left( \tau = z_{eff} \cdot \frac{T_e}{T_i} \right)$. In the following sections, a convergence test is presented, followed by heat loss from ETG turbulence at several locations in the pedestal. The dependence of heat loss from ETG turbulence on $\tau$ and $\eta_e$ is also discussed.

**ETG convergence test**

| nz | edge_opt | Other params | $Q_{es}A$ (MW) |
|---|---|---|---|
| 384 | 2 | | 1.34 |
| 384 | 4 | | 1.33 |
| 384 | 6 | | 1.37 |
| 512 | 2 | | 1.44 |
| 512 | 6 | nky*2, ky/2 | 1.52 |
| 512 | 6 | nx*1.5, lx*1.5 | 1.57 |
| 768 | 6 | | 1.4 |
| 1024 | 2 | | 1.55 |
| 1024 | 6 | | 1.47 |
| 1248 | 6 | | 1.55 |

Table 7: Nonlinear ETG simulation convergence test.

We notice that the mode structure of nonlinear ETG simulations of I-mode pedestal, eigenfunctions develop fine structure in the z dimension (slab-like), similar to observations in Refs. 17,18,19. We run a series of simulations increasing the number of grid points in the z dimension to find out an optimum number to use. Additionally, we tried to use different values of the *edge_opt* parameter to adjust the locations of grid point in z – the higher the value, the more crowded grid points are at the outboard midplane. The results are in Table 7.

In terms of resolution in other dimensions, we used $k_{ymin}\rho_s = 5$, with 48 positive $k_y$ Fourier modes, and radial box size $L_x = 2.9\rho_s$, and 128 positive and negative $k_x$ Fourier modes as the base case parameters. To test y dimension parameters, one simulation with $k_y\rho_s = 2.5$, nky = 96 is run for nz = 512 without finding a significant change in total heat loss. To test x dimension parameters, one simulation with nx = 192 and lx = 4.5$\rho_s$ is run and it also produced similar amount of total heat loss compared to the base case.

Based on these convergence tests, we choose to use nz = 512, kymin = 2.5, nky = 72 for the nonlinear ETG runs at different radial locations.

**Radial profile of parameters and heat loss**

The local flux tube nonlinear ETG simulations are conducted for 4 chosen radial locations: $\rho_t = 0.97$ is at the top of the electron temperature (Te) pedestal; $\rho_t = 0.975$ is where we see ITG/Impurity mode to peak for WCM related toroidal mode numbers; and $\rho_t = 0.985$ is about where the middle of electron pressure pedestal is.

We use Zeff = 2.8 to start with for all radial positions. Since Ti / Te >~ 0.8 for $\rho_t = 0.97$, 0.975 and 0.98, the parameter $\tau$ is as high as $\tau \sim 3.4$ for these locations.

As is shown in Table 8, The ETG caused heat loss at $\rho_t = 0.98$ and 0.985 is, respectively, 3.7 MW and 7.2 MW. These values of heat loss are close to (or higher than) the input power through the pedestal ($\sim 4$ MW).

| $\rho_t$ | $\hat{s}$ | $\eta e$ | $\tau$ | QesA (MW) |
|---|---|---|---|---|
| 0.97 | 4.23 | 1.74 | 3.38 | 0.1 |
| 0.975 | 3.58 | 2.86 | 3.45 | 1.0 |
| 0.98 | 3.42 | 4.32 | 3.30 | 3.7 |
| 0.985 | 4.99 | 5.75 | 2.88 | 7.2 |

Table 8: Radial location scan of heat loss from nonlinear ETG simulations.

**τ dependence**

As mentioned above, we define the parameter $\tau = z_{eff} \cdot \frac{T_e}{T_i}$, which determines the strength of the adiabatic ion response in ETG simulations. The two factors making up this parameter are both hard to measure accurately in the experiments. We therefore run a set of cases at $\rho_t = 0.975$ to determine the τ dependence of the total heat loss; the heat loss goes up from 1.0MW to 1.26 MW when τ goes down from 3.45 to 1.0 (Table 9).

| τ | $Q_{es}A$ (MW) |
|---|---|
| 1. | 1.26 |
| 2.8 | 1.11 |
| 3.45 | 1.0 |

Table 9: τ scan of heat loss from nonlinear ETG simulations at $\rho_t = 0.975$.

**$\eta_e$ dependency**

| $\rho_t$ | $\eta_e$ | $Q_{es}A$ (MW) |
|---|---|---|
| 0.97 | 1.74 | 0.1 |
| 0.97 | 2.09 | 0.3 |

| 0.975 | 3.58 | 1.0 |
|-------|------|-----|
| 0.975 | 3.43 | 1.8 |

Table 10: $\eta_e$ scan of heat loss from nonlinear ETG simulations at $\rho_t = 0.97$ and 0.975.

We increase $\eta_e$ by 20% for the two radial locations, $\rho_t = 0.97$ ($\rho_t = 0.975$), where nominal ETG nonlinear heat loss is enough (a little lower than) to satisfy power balance. The heat loss is boosted up 3 (2) times the nominal $\eta_e$ value. This ETG heat transport in the pedestal is, therefore, stiff and is in the right range to match power balance (with a minor contribution from the ITG/impurity mode).

In summary, ETG turbulent transport can, in principle, match power balance for this shot in the middle of the pedestal (Fig 12).

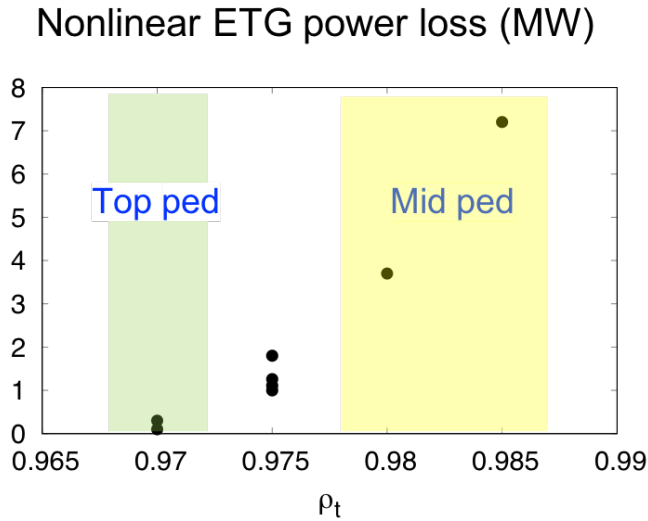

Figure 12: Nonlinear power loss (MW) from ETG turbulence. Results from $\eta_e$ scan and $\tau$ scan are also added to the graph.

**IMPURITY CONFINEMENT TIME**

High Z impurity laser blow-off experiments[13] show that the core impurity transport is anomalous ($D_z^{eff} >> D_{nc}$) for I-modes. Experiments also find the following scalings of the impurity confinement time ($\tau_Z$) with plasma parameters: $\tau_Z$ increases with plasma current and decreases with input ICRF power. The I-mode case we simulated, has $I_p = 1.2$ MA, and $P_{ICRF} = 5$ MW. The experimental point in Fig 14 of Ref 10, closest to it (our I-mode case has higher power and slightly higher current), has $\tau_z \sim 30$ ms. The experimental scaling, then, will predict that these two effects are likely to roughly cancel. Therefore, we estimate the empirical impurity confinement time to be $\tau_z \sim 30$ ms. (Note that this is on the same scale of energy confinement time[10].)

In the impurity injection experiments, one encounters two types of impurities: the intrinsic population, and externally injected impurities from the laser blow-off. The average charge of the intrinsic impurities is estimated to be $Z_{ave} = 10$. The laser injected Calcium happens to have about the same charge for pedestal temperatures in this shot. Hence our pedestal simulations assume only a single impurity species of $Z = 10$. We model the impurity injection experiments as follows: for $t < 0$ we assume that the pedestal plasma is at a steady state with no impurity flux; at $t = 0$, the impurities are injected to raise the impurity density at the top of pedestal and inside in the plasma core. The experiments observe that the impurity profile is flat in the core[16] and we assume that as well. However, the impurity density at the separatrix is assumed to be unperturbed during the decay phase of impurity injection. This is reasonable since impurity life time in the SOL is extremely short and the SOL impurity density is determined by the balance of rapid loses due to parallel motion and impurities source due to sputtering from plasma facing components. These processes are extremely complex and effectively impossible to calculate accurately. However, it is reasonable to assume that the SOL impurity density is essentially unperturbed by the laser blow-off since the impurities introduced by laser will be very rapidly flushed out from the SOL. Hence the ambient impurity density in SOL will not be strongly affected.

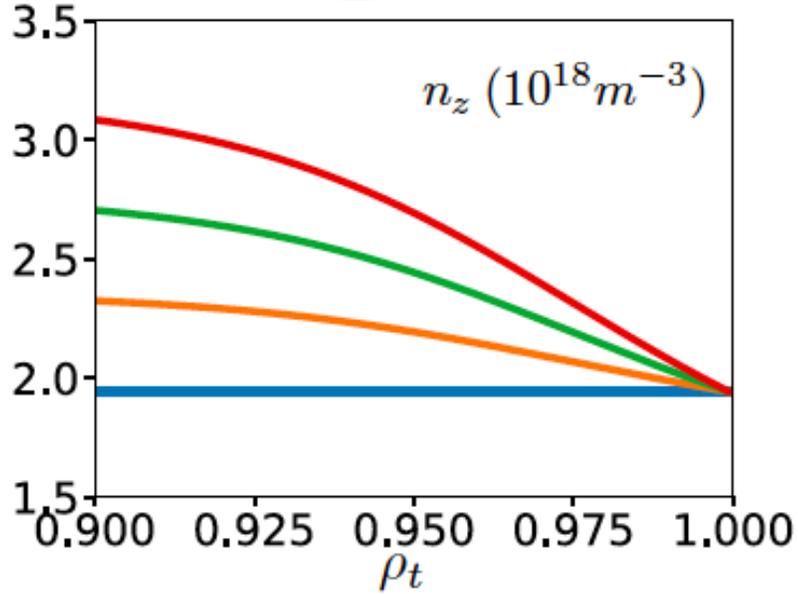

Figure 13: Impurity profiles in the pedestal for the simulations designed to estimate impurity confinement time.

We designed a set of impurity profiles (Fig 13) to reproduce the decay phase of impurity injection in laser blow-off experiments. It is assumed that the impurity density at $\rho_t$=1 is same for all. The effect of impurity laser blow-off, therefore, is to boost up the impurity density gradient in the pedestal $n_{z,\,ped}$ since the impurity density in the pedestal (and inside) increases while the separatix density remains essentially unperturbed. We use nonlinear simulations, then, to calculate the increased impurity flux from this boosted gradient and go on to calculate the core impurity decay rate for comparison with experimentally measured decay rate.

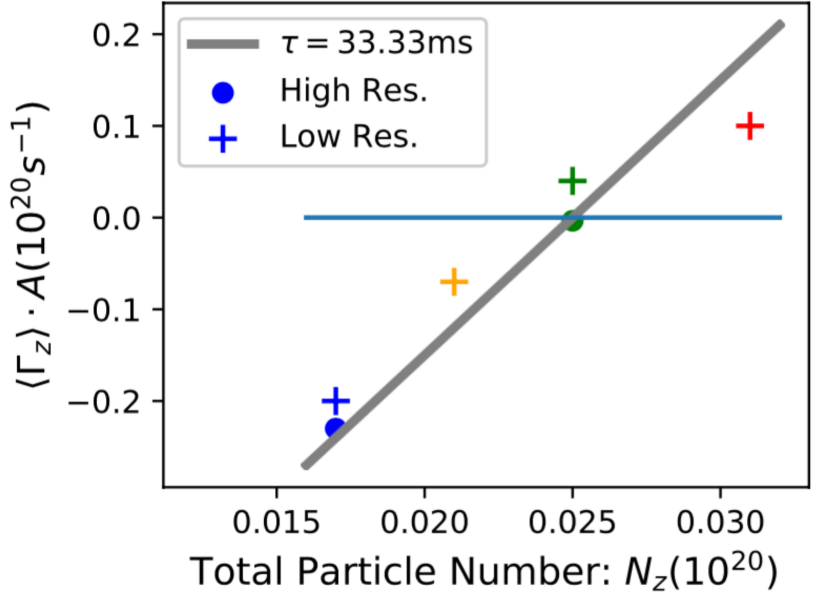

Figure 14: Plot of impurity particle loss rate versus total impurity particle number. The colors of the symbols correspond to those of the profiles shown in Fig. 13. The slope represents the inverse of the impurity confinement time. A line representing τ=33 ms is shown for reference. High resolution simulations (n=3-189 with hyperdiffusion) are shown with solid circles and lower-resolution simulations (n=4-60 without hyperdiffusion) are shown with plus symbols, demonstrating that the prediction of impurity confinement time is insensitive to the details of the $k_y$ spectrum and, presumably, cross-scale interaction.

Impurity particle losses from global nonlinear simulations with these profiles are used to compute the impurity confinement time from ITG/impurity modes, as follows:

Because the impurity profile is flat in the core, the total number of impurities in the plasma, the inventory, $N_z$ is about plasma volume V times the impurity density at the top of the pedestal $n_{z, ped}$.

The rate of change of impurity inventory is determined by the impurity flux from the pedestal:

$$\frac{d\,\Delta N_z}{d\,t} = <\Gamma_z> \cdot A$$

where $<\Gamma_z>$ is the surface averaged impurity flux. If the RHS is a linear function of $N_z$, then this equation describes an exponential decay to a steady state constant value. The steady state value is when the RHS =0, as indicated above. The decay rate can be found by plotting the RHS against $N_z$ and examining the slope of the line and shown in Fig 14. We can estimate (from the inverse slope) the impurity confinement time: $\tau_z \sim 33$ ms; this value is quite close to the experimental value.

### SYNTHETIC DIAGNOSTICS AND GEODESIC ACOUSTIC MODES

Geodesic-acoustic modes (GAM) and WCM are observed in I-mode on C-Mod using Gas-Puff-Imaging (GPI) based diagnostics[20], which mainly measure density fluctuations and can be analyzed to determine oscillations in poloidal velocity $v_{pol}$. The pattern of I-modes is similar: there is a spectral peak corresponding to a WCM, and an even stronger peak, seen most clearly in $V_{pol}$ near $k_\theta \sim 0$ with frequency $\sim (2T_e/m_i)^{1/2}/R$, which is consistent with Geodesic Acoustic Modes (GAMs). GAMs are measured to exist approximately at the same location as the Er-well and WCM and interact with WCM[20].

We construct a synthetic diagnostic tool to qualitatively mimic GPI to see if our GENE simulations give similar results. There was no GPI measurement for our particular shot, so we can only expect qualitative agreement with GPI results from other I-mode shots.

Fig 9(a) in Ref. 11 shows the emission fluctuations measured by GPI at the horizontal midplane during an I-mode phase of C-Mod discharge 1100204022 (1.3 MA, 5.8 T, upper single null). A weakly coherent mode is visible, centered at about $f \sim 220$ kHz, $k_\theta \sim 1.25$ cm$^{-1}$ (electron diamagnetic direction). There's also even higher intensity at $k_\theta \sim 0$ and frequency $\sim 20$–50 kHz, which are GAMs.

To construct a synthetic diagnostic, we first take note of the character of GPI. It takes pictures on a 2D array and the third dimension is integrated over the line of sight. Due to misalignment of its sightline with the local magnetic field line ($\theta_{mis} \sim 6°$) as well as variation of magnetic field line pitch angle within the gas cloud, there's a finite spatial resolution of the GPI diagnostics[21] and higher wavenumber fluctuations along the line of sight are averaged out.

We implement our synthetic diagnostic for GPI to be convenient for GENE coordinates. The fluctuations we get from GENE simulations are expressed as

$$f \sim f(\theta, r)e^{i\,n\,(\zeta - q\,\theta)}$$

where θ and ζ are the magnetic coordinates for poloidal and toroidal angle and r is the radial coordinate. The coordinates and representation used in GENE are chosen so that the scale of variation of θ is on the order of the parallel correlation length. Rapid variations on the scale of $k_\perp$ are due to the eikonal. The GPI image is in the (r, θ) plane averaging over the line of sight. This requires us to find an appropriate way to average over ζ. To simplify the problem, we Taylor expand the magnetic angle coordinates in the small gas cloud volume, which is a good approximation since the gas cloud dimensions is small compared to the magnetic equilibrium scales. We consider the origins of coordinates for all spatial distances to be at the center of the gas cloud. We expand $(\zeta - q\,\theta)$, obtaining $(\zeta - \frac{B_{tor}}{B_{pol}} \cdot \frac{y}{R})$, where y is the distance in the poloidal direction tangential to a flux surface. We then put this in terms of the toroidal distance $(l)$, which is nearly the same as the sightline distance: $(l - \frac{B_{tor}}{B_{pol}} \cdot y)/R$. For convenience, we define the pitch angle of magnetic field line to be $tan\ tan\ (\theta_B) = \frac{B_{tor}}{B_{pol}}$. The GPI averages over the line of sight, which we approximate to be given by the line $(\theta_{sight}) \cdot l$. For every toroidal mode number n, we integrate along the sightline the assumed gas cloud emissivity profile:

$$I(l) \sim \frac{1}{r^2 + l^2}$$

where r is the perpendicular distance from gas nozzle to the observation point], and include the eikonal phase variation:

$$\int\ \ dl\ I(l)\ e^{in\zeta}$$

Then the final synthesized density perturbation, as it would appear in the 2D GPI image, is

$$n(r, \theta) = \Sigma I_n n_n(r, \theta)\ e^{-inq\theta}$$

where $I_n = e^{-n\frac{r}{R}\frac{|\theta_{sight} - \theta_B|}{\theta_B}}$, where θ_sight - θ_B is the mis-alignment angle between light of sight and true magnetic field direction. (We have also approximated tan θ ~ θ, since all the angles are small.) Notice if there's no misalignment, θ_sight - θ_B = 0, then there's no reduction of resolution via the reduction of higher wavenumbers, since all I_n = 1.

This result is approximately what the GPI would observe by averaging over the sightline. Note that an additional level of signal processing, Velocimetry, is usually implanted for the GPI image of density fluctuations, to give the velocity fluctuation which is the intrinsic signature of a GAM and not just the density fluctuation; we have not implemented this.

The density fluctuation spectra in the nonlinear saturation state under different simulation conditions are not the same. For one simulation shown in Fig 15(a), we were able to reproduce a qualitatively similar spectrum to the experimentally observed ones, with both WCM and GAM, at about the frequencies expected. The peak of the frequency is around 200—250 kHz which is slightly lower than the WCM (~ 300—400 kHz), as measured for this shot. The GAM frequency is ~ 20–50 kHz, as expected. As can be seen from our spectrum, there's a strong fluctuation in that range at ky =0. For other simulations (an example shown in Fig 15(b)), we see a broad band of unstable modes each with comparable density fluctuation level. The same procedure of GPI filtering applied to those runs doesn't give us a clear peak to match the observed WCM. Mismatch between the density fluctuations frequency spectrum and that of the observed WCM could arise from box parameters in our simulations, and experimental error in profiles. We leave more systematic sensitivity test to future work.

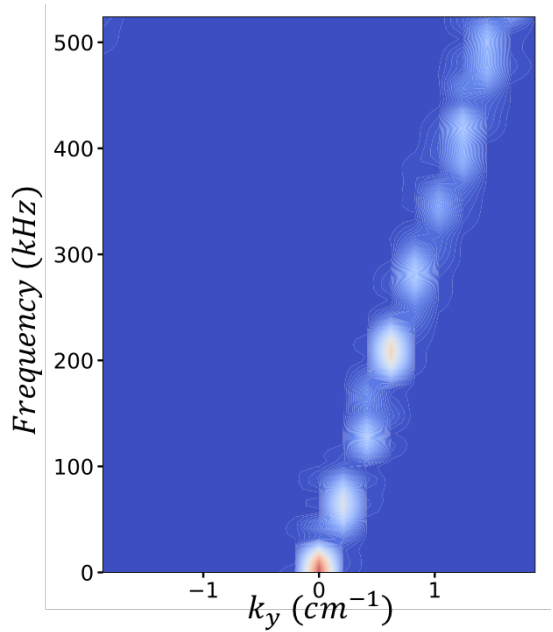

Figure 15(a): An example of results of GPI synthetic diagnostic that has some characteristics similar to observations on the referenced I-mode case.

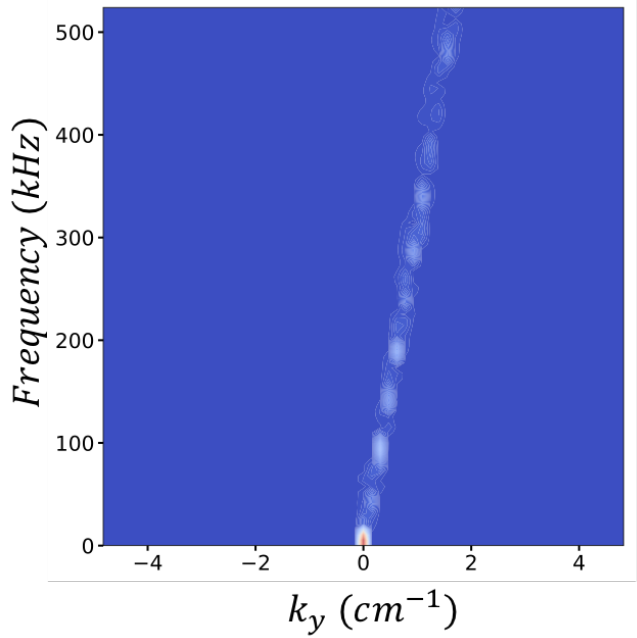

Figure 15(b): An example of results of GPI synthetic diagnostic that doesn't match experiment.

A well-known feature of fluctuation spectra of C-Mod I-mode is the strong fluctuations of GAMs. We note the features of this slab-like ITG/Impurity mode, the frequency, growth rate and parallel wavenumber, are such that the nonlinear beating of two such modes has the frequency and parallel wavenumber of GAMs. Therefore, WCMs, if they were slab-like ITG/Impurity mode fluctuations, would likely lead to beat fluctuations which are similar to fluctuations observed as GAMs.

## Discussions and Conclusion

In a pedestal where temperature, electron density and impurity density profiles are all in steady state, such as I-mode, transport must arise in all these channels to enforce the steady state. Our simulations indicate that there are two different modes that are primarily responsible for transport in different channels. Most energy losses are due to ETG modes. However, impurity particle transport and probably electron particle transport are due to the WCM-like mode: our simulations found that WCM is a slab-like ITG/Impurity mode. These two modes are consistent with estimates of inferred diffusivities. In particular, the impurity diffusivity is on the order of the total thermal diffusivity ($D_Z \sim \chi_e$) and the main electron diffusivity is much smaller than the electron thermal diffusivity ($D_e \ll \chi_e$). The fingerprints of the ETG mode imply that it causes, almost exclusively, transport in the electron particle channel. In the parameter regime of C-Mod I-mode here, the slab-like ITG/Impurity mode causes primarily impurity particle diffusivity, ion thermal diffusivity, but little electron particle diffusivity. Since both passing and trapped electron are almost adiabatic, it follows that the electron particle transport is weak for this mode.

This conclusion results from multiple parameter scans in the simulation. Variations of ion temperature and density gradients, electron density gradient, impurity gradient, and collisionalities indicate that this mode behaves like a slab-like ITG/impurity mode and is inconsistent with a Resistive Ballooning Mode. In addition, nonlinear simulations of the ITG/Impurity mode show the presence of large amplitude of fluctuations with the characteristics of GAMs qualitatively similar to experiment. The linear and nonlinear simulations find that the electron particle transport is much less for this mode.

Impurity outflux computed from the nonlinear simulations of the ITG/Impurity modes indicates that this mode could cause strong impurity particle transport. The experimental impurity confinement time measured by the laser blow-off experiment is reproduced by nonlinear simulations of impurity profile variations.

We examined the sensitivity of power loss through ETG turbulence on profile gradients and the impurity level. Nonlinear simulations show that the ETG turbulence could be responsible for dominant heat transport by matching experimental power balance in the middle of the pedestal.

In summary, simulation result of ITG/Impurity modes and ETG modes are consistent with multiple experimental observations. Going forward, we think it would be particularly useful to obtain a better estimate of the experimental particle source. This would allow a better-inferred electron particle diffusivity $D_e$. A distinguishing feature of the results here is a low value of $D_e/\chi$ and $D_e/D_z$. This differentiates our work from the conclusion of others that the WCM is a Resistive Ballooning mode.

In addition, future simulations could consider I-mode-like profiles for burning plasma parameters to see if the modes identified here are capable of sustaining steady state ELM-free scenarios with good energy confinement.

## Acknowledgements

This material is based upon work supported by the U.S. Department of Energy grant DE-FG02-04ER54742, and user Facility Alcator C-Mod award DE-FC02-99ER54512, the National Energy Research Scientific Computing Center and the Texas Advanced Computing Center.